\documentclass[pre,preprint,showpacs]{revtex4}
\usepackage{epsfig}
\usepackage{amssymb}
\begin{document}

\title{Nonlinear stage of the Benjamin-Feir instability:
Three-dimensional coherent structures and rogue waves
} 
\author{V.~P. Ruban}
\email{ruban@itp.ac.ru}
\affiliation{Landau Institute for Theoretical Physics,
2 Kosygin Street, 119334 Moscow, Russia} 

\date{\today}

\begin{abstract}
A specific, genuinely three-dimensional mechanism of rogue wave formation, 
in a late stage of the modulational instability of a perturbed 
Stokes deep-water wave, is recognized through numerical experiments. 
The simulations are based on fully nonlinear equations 
describing weakly three-dimensional potential flows of an ideal fluid with a 
free surface in terms of conformal variables.
Spontaneous formation of zigzag patterns for wave amplitude is observed  
in a nonlinear stage of the instability. If initial wave steepness 
is sufficiently high ($ka>0.06$), these coherent structures 
produce rogue waves. The most tall waves appear in ``turns'' of the zigzags.
For $ka<0.06$, the structures decay typically without formation 
of steep waves. 

\end{abstract}




\maketitle


The problem of rare extreme localized sea surface excitations 
(known as freak waves, rogue waves, or killer waves) 
has been extensively studied for last years (see, e.g., review 
\cite{Kharif-Pelinovsky}, where different physical mechanisms 
of the rogue wave phenomenon are discussed; for recent results in this field,
see Ref.~\cite{RogueWaves2006}, and references therein). 
It has been recognized that one of the most important reasons for freak 
waves is the Benjamin-Feir (modulational) instability of a 
plane propagating wave \cite{Benjamin-Feir,Zakharov67}.
In the most probable scenario, refraction of swell in a spatially non-uniform 
current causes preliminary amplification of wave height around caustic region 
\cite{Peregrine,LP2006}. As a result, self-attractive nonlinear interaction
becomes strong, so freak waves develop. 
Several simplified weakly nonlinear models, in particular 
the nonlinear Schroedinger equation and its generalizations
\cite{Dysthe1979,TKDV2000,OOS2000,Janssen2003,OOS2006,SKEMS2006}, 
have been suggested to describe rogue waves quantitatively.
But recently it has been realized that even the fully dispersive
Zakharov equations, the most accurate among weakly nonlinear models, 
are still poor for describing rogue waves in their final stage, 
when the wave steepness is large \cite{LZ2005}. It should be noted that
a peak-to-trough height $h$ of a rogue wave reaches values which are typical for
a corresponding limiting Stokes wave, that is $0.10<h/\lambda \lesssim 0.14$,
where $\lambda$ is a typical wave length. In the absence of 
an accurate analytical theory, there is a strong need in numerical results 
for fully nonlinear water waves. 
Recently, very precise simulations of exact equations for two-dimensional 
(2D) flows of an ideal fluid with a free boundary   
were reported and confirmed appearance of rogue waves in a nonlinear stage of
the modulational instability \cite{CFGK2006,ZDP2006,DZ2005Pisma,ZDV2002}.
Many of those results are based on a compact (1+1)-dimensional form of 
governing equations in terms of so called conformal variables
\cite{DKSZ96,DZK96,D2001,R2004PRE,R2005PLA}, when a shape of the boundary
is represented parametrically as $x=u-\hat H Y(u,t)$, $y=Y(u,t)$, where
$x$ is a horizontal coordinate, $y$ is a vertical coordinate, $\hat H$ is 
the Hilbert operator, and $Y(u,t)$ is an unknown real-valued function.
The second unknown function is a boundary value $\psi(u,t)$ of the 
velocity potential. It is important that complex function 
$Z(u,t)=u+(i-\hat H)Y(u,t)$ is an analytical function $z(w,t)$
evaluated at the real axis of the complex variable $w=u+iv$. Function 
$z(w,t)$ corresponds to a conformal map of the lower half-plane of
variable $w$ onto the flow region in complex plane $z=x+iy$.
This description effectively ``straightens'' the free boundary, 
and it makes possible simulations of planar potential flows on a 
personal computer with a spatial resolution up to several million points. 
However, real flows are always three-dimensional (3D), 
and the third spatial dimension definitely plays an important role 
in the wave dynamics. Unfortunately, the only ``exact'' numerical method 
for 3D potential flows -- the boundary element method -- is rather
``expensive'' despite its various sophisticated modifications, 
so the best spatial resolution available now is limited to 
a few thousand points representing a free surface 
\cite{Clamond-Grue-2001,Fructus_et_al_2005,Grilli,
GuyenneGrilli2006,FochesatoDias2006,DiasBridges2006}.
But the problem of rogue waves requires much larger computational domains,
at least several million points. 
On the other hand, since the modulational instability is predominantly 
two-dimensional, typical rogue waves appear long-crested.
For long-crested waves,  fully nonlinear approximate equations
in conformal variables have been derived recently \cite{R2005PRE,RD2005PRE}.
The dynamical variables $Y(u,q,t)$ and $\psi(u,q,t)$ now slowly vary along
the second horizontal coordinate $q$. The corresponding weakly 3D theory 
is based on a geometric small parameter, the ratio of a typical wave length 
$\lambda$ to a large scale $l_q$ along wave crests. 
Most recent numerical results based on this theory correspond to a spatial
resolution about 16 million points \cite{R2006PRE}. 
Important 3D effects have been observed.
In particular, the above cited work reports observation of
long-lived oscillating rogue waves. It has also  been demonstrated there, 
for one example, that in evolution of a randomly perturbed planar wave,  
specific coherent zigzag structures are formed, which consist of obliquely 
oriented stripes with heightened wave amplitude.

\begin{figure}
\begin{center}
(a)\epsfig{file=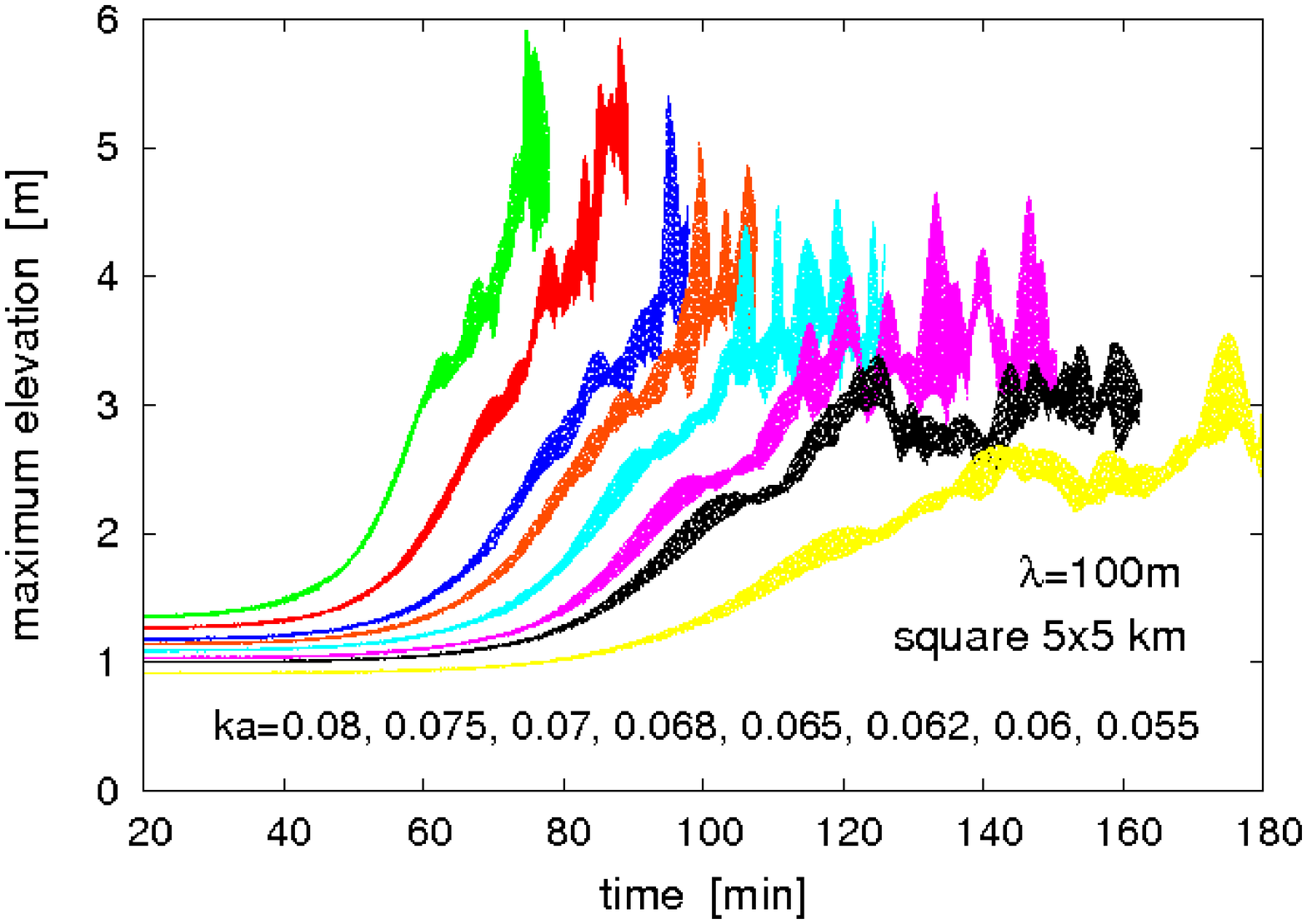,width=125mm}\\
(b)\epsfig{file=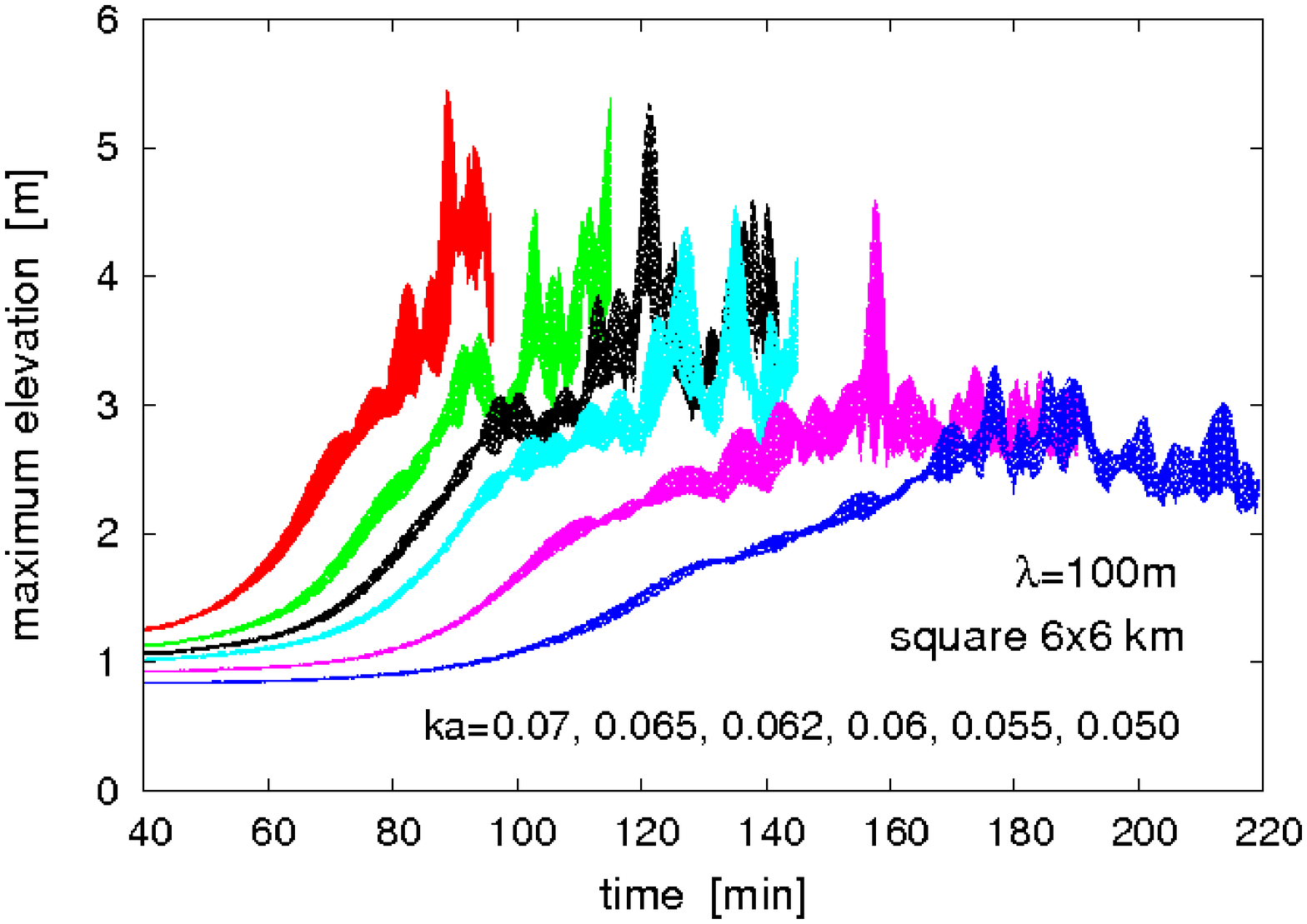,width=125mm}     
\end{center}
\caption{(Color online). Maximum elevation of the free surface versus time in 
two sets of numerical experiments with a randomly perturbed planar wave of 
$\lambda=2\pi/k=100$ m, for different initial amplitudes $a$ of the first 
Fourier harmonics. 
Oscillations with a typical period about 16 s are not resolved in this figure.} 
\label{Y_max-I-II} 
\end{figure}

In the present work, 3D aspects of the Benjamin-Feir instability are considered
in more detail. Late stages of the instability are simulated
for different amplitudes of the initial plane wave, 
and then a comparison is made.
The most important results of these numerical experiments are the following: 
 (i) a new, genuinely 3D mechanism of freak wave formation 
has been recognized, when an extreme wave appears in a ``turn'' of a 
zigzag-shaped stripe of an increased wave amplitude;
 (ii) for the first time,  two different dynamical regimes in a nonlinear stage 
of the Benjamin-Feir instability in 3D space have been observed. 
Let $k$ be a wave number, and let $a$ be an amplitude of the first 
Fourier harmonics of a function $Y_0(u)$ representing the initial wave.
If $ka >0.06$ (that corresponds to $h_0/\lambda \ge 0.02$), 
then after an initial stage of the Benjamin-Feir instability,
rogue waves with a ratio $h/\lambda \gtrsim 0.08$ are frequent in a 
wave system of a size $L\sim 50 \lambda$. 
(see Fig.\ref{Y_max-I-II}). 
On the contrary, for $ka\le 0.06$, the probability to observe a tall wave 
with $h/\lambda \gtrsim 0.07$ in a finite-size domain is very small.
Existence of such a ``threshold'' is an important, essentially 3D effect.

{\it Equations of motion.---}
The nonlinear theory for surface waves on infinitely deep water, 
developed in Refs.~\cite{R2005PRE,RD2005PRE}, is used 
herein. Equations of motion for unknown functions can be written in a 
non-canonical Hamiltonian form, involving variational derivatives 
$(\delta{\cal K}/\delta\psi)$ and $({\delta{\cal K}}/{\delta Z})$, where
${\cal K}\{\psi,Z,\bar Z\}$ is the kinetic energy \cite{R2005PRE,RD2005PRE},
and $\bar Z$ is the complex conjugate of $Z(u,q,t)\equiv u+(i-\hat H)Y(u,q,t)$.
The Hilbert operator $\hat H$ is diagonal in the Fourier representation:
it multiplies the Fourier-harmonics 
$
Y_{km}(t)\equiv\int Y(u,q,t)e^{-iku-imq}du\,dq
$
by $[i\,\mbox{sign\,}k]$,  so that
\begin{equation}\label{HY}
\hat H Y(u,q,t)=\int [i\,\mbox{sign\,}k] Y_{km}(t)e^{iku+imq}
{dk\, dm}/({2\pi})^2.
\end{equation}
The first equation is the so called kinematic condition on the free surface:
\begin{equation}\label{kinematic}
Z_t=iZ_u(1+i\hat H )\left[\frac{(\delta{\cal K}/\delta\psi)}
{|Z_u|^2}\right].
\end{equation}
The second equation is the dynamic condition (Bernoulli equation):
\begin{eqnarray}
\psi_t&=&-g\,\mbox{Im\,}Z
-\psi_u\hat H\left[\frac{(\delta{\cal K}/\delta\psi)}{|Z_u|^2}\right]
\nonumber\\
\label{Bernoulli}
&+&\frac{\mbox{Im}\left((1-i\hat H)
\left[2({\delta{\cal K}}/{\delta Z})Z_u
+ ({\delta{\cal K}}/{\delta\psi})\psi_u\right]\right)}{|Z_u|^2},
\end{eqnarray}
where $g$ is the gravitational acceleration.
Equations (\ref{kinematic}) and (\ref{Bernoulli}) completely determine
evolution of the system, provided the kinetic energy functional 
${\cal K}\{\psi,Z,\bar Z\}$ is explicitly given. 
Unfortunately, in three dimensions there is no exact compact expression 
for ${\cal K}\{\psi,Z,\bar Z\}$. However,
for long-crested waves propagating mainly in $x$ direction 
(the parameter $\epsilon\sim(\lambda/l_q)^2\ll 1$),
we have an approximate kinetic-energy functional in the form
\begin{equation}\label{K_approx}
{\cal K}\approx
\tilde{\cal K}={\cal K}_0+\tilde{\cal F}=
-\frac{1}{2}\int \psi\hat H\psi_u \,du\,dq +\tilde{\cal F},
\end{equation}
where the first term ${\cal K}_0$ describes purely 2D flows, 
while weak 3D corrections are introduced by the functional $\tilde{\cal F}$:
\begin{eqnarray}
\tilde{\cal F}&=&\frac{i}{8}
\int(Z_u\Psi_q-Z_q\Psi_u)\hat G
\overline{(Z_u\Psi_q-Z_q\Psi_u)}\,du\,dq
\nonumber\\
&+&\frac{i}{16}\int\Big\{\left[
(Z_u\Psi_q-Z_q\Psi_u)^2/{Z_u}\right]\,\hat E\overline{(Z-u)} 
\nonumber\\
&&\quad - (Z-u)\,\hat E\overline{\left[(Z_u\Psi_q-Z_q\Psi_u)^2/{Z_u}\right]}
\Big\}\,du\,dq.
\label{H_modified}
\end{eqnarray}
Here $\Psi\equiv (1+i\hat H)\psi$, and
the operators $\hat G$ and  $\hat E$ are diagonal in the Fourier
representation:
\begin{equation}\label{GE_def}
G(k,m)=\frac{-2i}{\sqrt{k^2+m^2}+|k|},\quad
E(k,m)=\frac{2|k|}{\sqrt{k^2+m^2}+|k|}.
\end{equation}
A difference between the above expression (\ref{K_approx})
and the unknown true water-wave kinetic energy functional is of order 
$\epsilon^2$, since $G(k,0)=1/(ik)$ for positive $k$, and $E(k,0)=1$
(see Refs.~\cite{R2005PRE, RD2005PRE}).
Besides that, the linear dispersion relation resulting from 
$\tilde{\cal K}$ is correct in the entire Fourier plane. 
It should be noted that in Ref.~\cite{RD2005PRE} a different
expression for $\tilde{\cal K}$ was used, with the same properties. 
However, if the form of ${\cal K}_0$ is fixed as given above, 
then the function $G(k,m)$ is determined uniquely. Function $E(k,m)$
was taken satisfying the relation $E(k,m)/G(k,m)=ik$ for $k>0$.

\begin{figure}
\begin{center}
   \epsfig{file=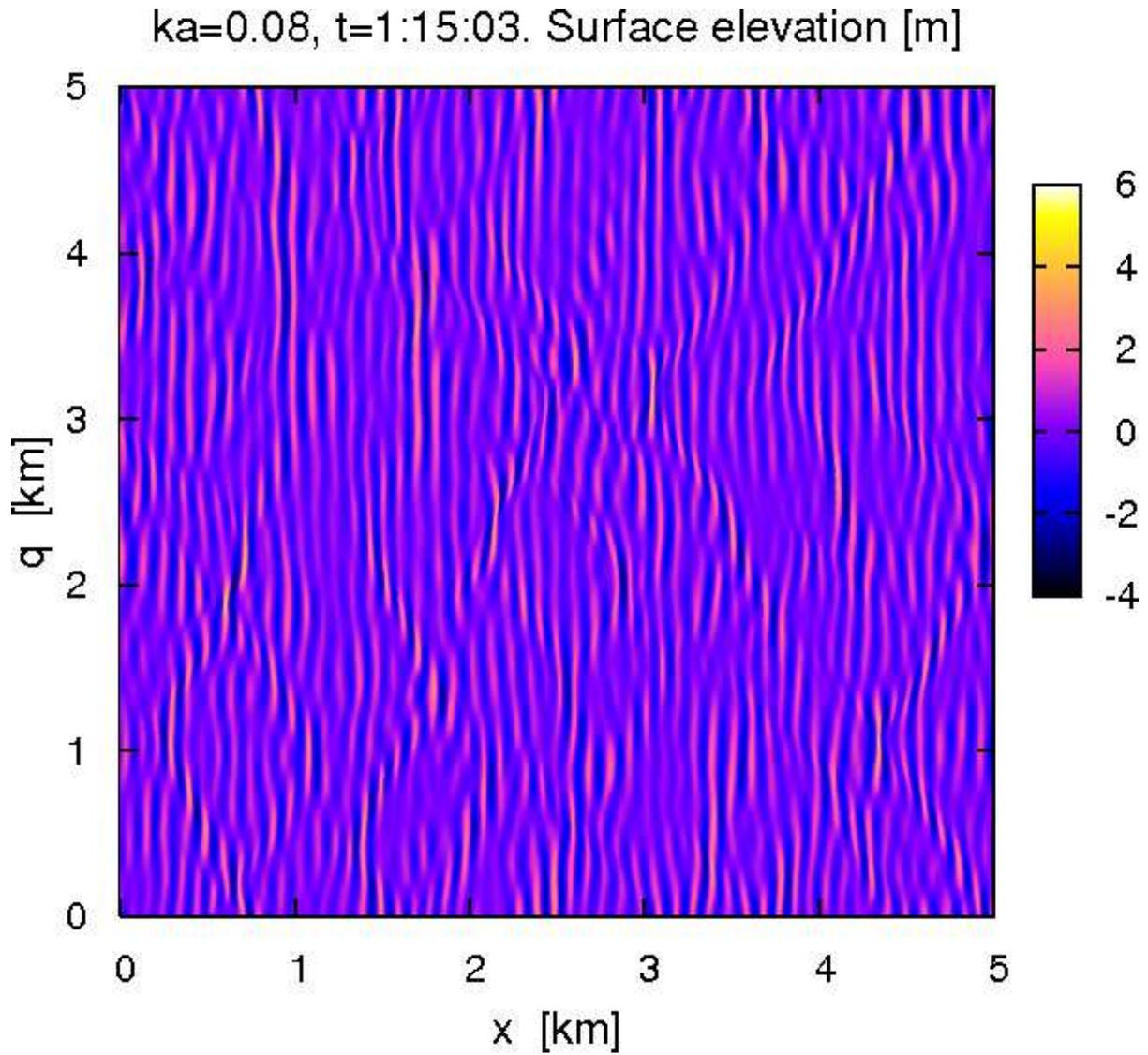,width=160mm}  
\end{center}
\caption{(Color online). Result of the Benjamin-Feir instability for 
$ka=0.08$, t=75 min 3 s. Zigzag coherent structures 
and several big waves are seen. Here a characteristic value of the parameter 
$\epsilon$ is approximately $0.15$.} 
\label{Ymap_ka008_75min} 
\end{figure}

\begin{figure}
\begin{center}
   \epsfig{file=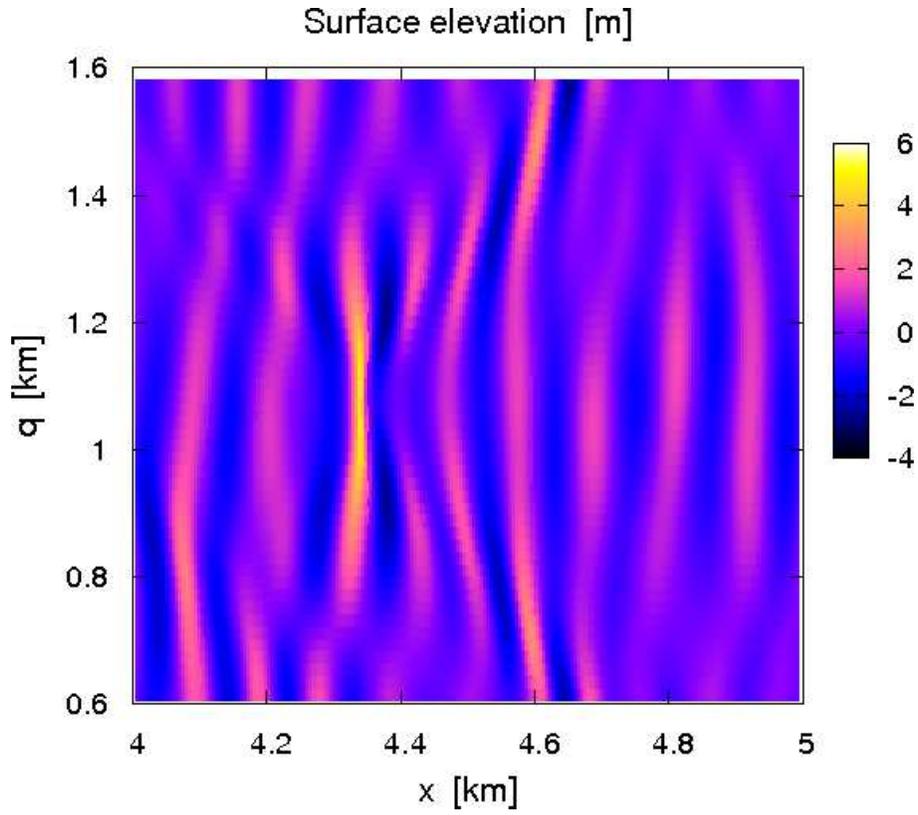,width=120mm}  
\end{center}
\caption{(Color online). 
Sub-region of Fig.\ref{Ymap_ka008_75min} with a rogue wave
at a zigzag turn. Wave crests in front of the big wave are obliquely
oriented in a focusing manner.} 
\label{zoom} 
\end{figure}

\begin{figure}
\begin{center}
   \epsfig{file=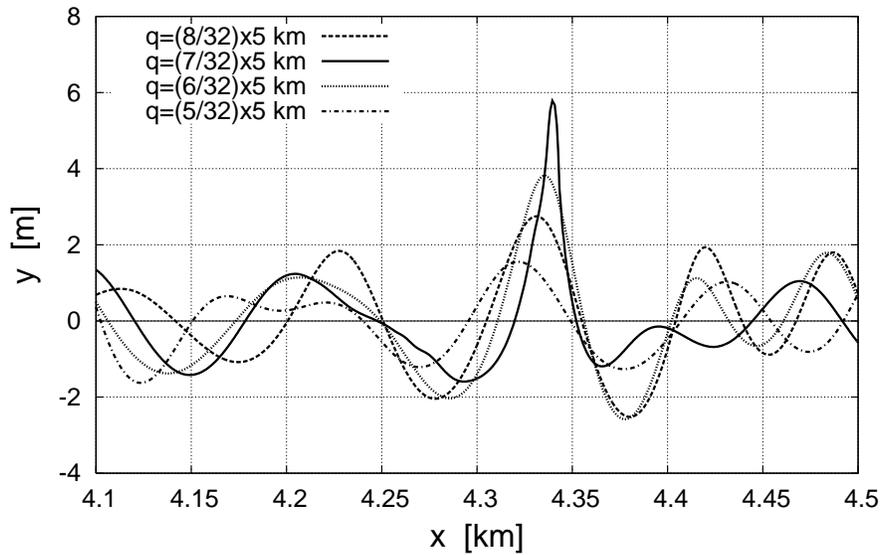,width=120mm}  
\end{center}
\caption{Profiles of the rogue wave from Fig.\ref{zoom}.} 
\label{Prof_ka008_75min} 
\end{figure}

{\it Numerical results and discussion.---}
The above equations were simulated for a randomly perturbed right-propagating
initial wave ($\lambda=100$ m) in two sets of numerical experiments.
The computational domain was a square $5\times 5$ km in the first set, and
$6\times 6$ km in the second set, with double-periodic boundary 
conditions. Since initial amplitudes were taken relatively small 
($ka\le 0.08$, see Fig.\ref{Y_max-I-II}), an exact stationary solution 
(Stokes wave) was approximated as follows: 
$kY_0(u)\approx -0.5 (ka)^2+ka\cos(ku)+(ka)^2\cos (2ku)+1.4(ka)^3\cos (3ku)$.
The same random-phase initial perturbation was used as in Ref.\cite{R2006PRE}.
The modulational instability developed initially in a similar way
for all the runs: an exponential growth of the perturbations, and then 
formation of coherent zigzag structures. At the end of this first stage, 
maximal amplitudes reached values approximately 
$2.5$ times exceeding the corresponding initial amplitudes 
(see Fig.\ref{Y_max-I-II}). However, a rather sharp difference 
in the dynamical behavior was observed in a later stage.
If $ka > 0.06$, then freak waves were intensively formed, 
typically in ``turns'' of the zigzags 
(see Figs.~\ref{Ymap_ka008_75min}-\ref{Prof_ka008_75min}). 
Amplitudes of the freak waves exceeded four times the corresponding initial 
amplitudes, and the steepness $\theta$ exceeded the value 
$\pi/6$ which is limiting for a Stokes wave (not shown). 
On the contrary, for $ka<0.06$, the waves remained quite smooth for 
most part of the time, and there was a tendency towards decay 
of the coherent structures.
It is reasonable to assume that this intermediate regime should bring the system
after some time to a weakly turbulent state, similar to what was observed in 
Refs.~\cite{OOSRPZB2002,DKZ2004}. 
However, our weakly 3D model is not applicable for the weak turbulence.

The above results imply that in a narrow parameter region 
$ka= 0.060\pm 0.005$, a relative amplification of the
initial plane wave is maximal, so freak waves are most prominent.
Of course, the ``threshold'' should be 
understood in a statistical sense, because details of the dynamics 
depend on a particular realization of the initial random perturbation.
Obviously, appearance of relatively high waves for $ka < 0.06$ cannot be 
excluded [see, e.g., the case $ka=0.055$ in Fig.\ref{Y_max-I-II}(b)],
but the corresponding probability is quite small, 
while at $ka > 0.06$ the probability to have a freak wave (in a finite domain
of a size about $50\lambda$) is comparable to 1.

\begin{figure}
\begin{center}
(a)\epsfig{file=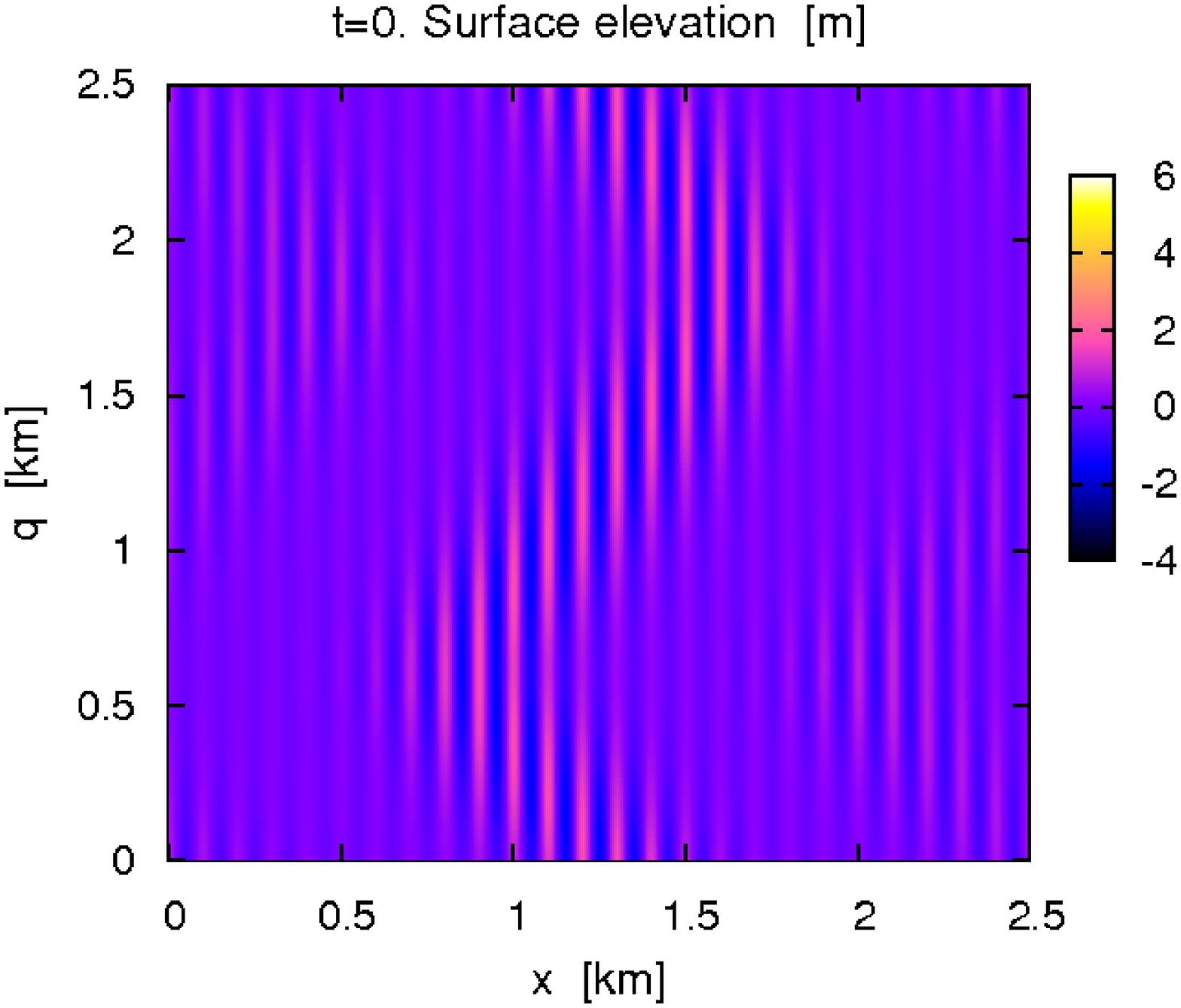,width=115mm}\\
(b)\epsfig{file=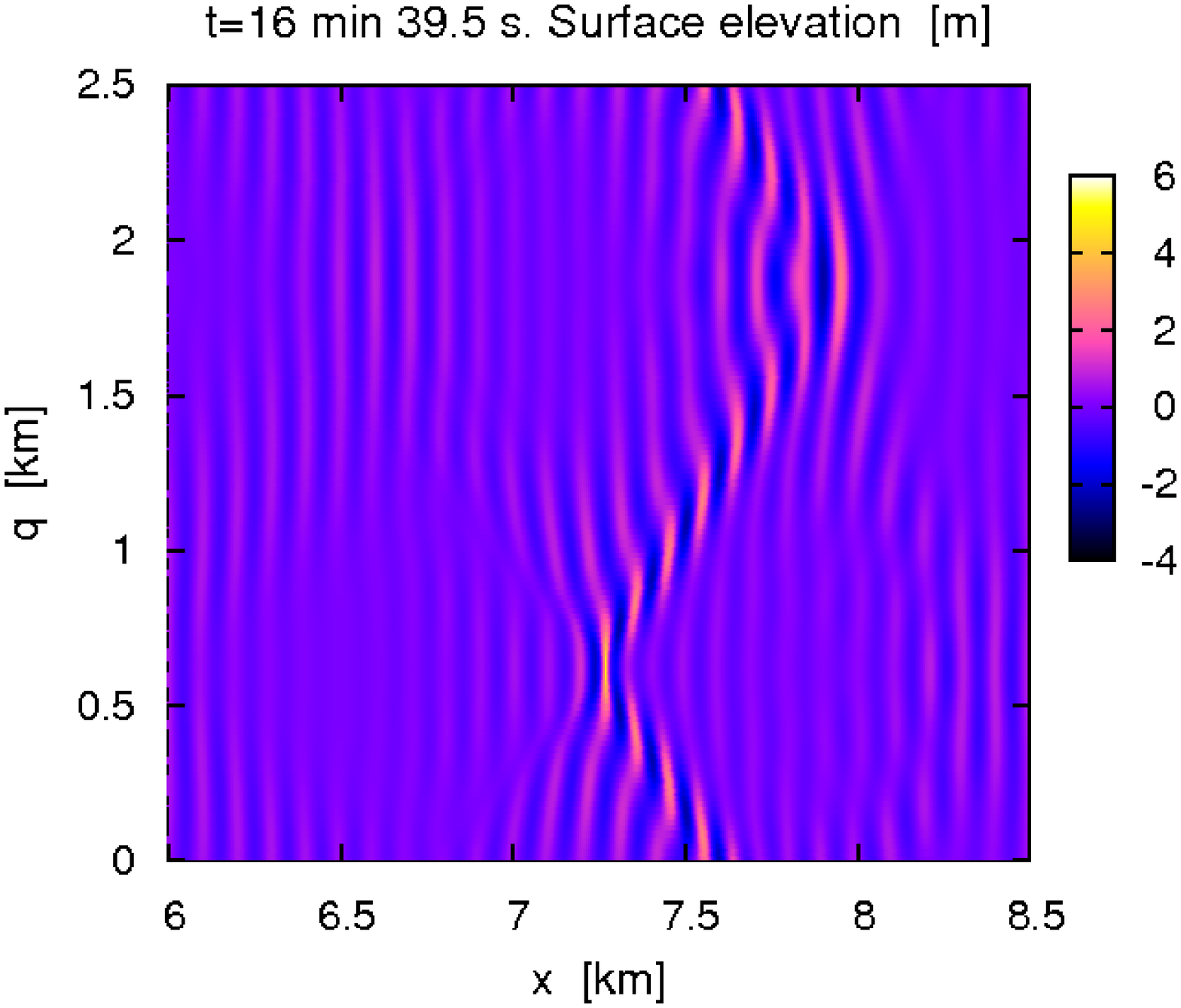,width=115mm}    
\end{center}
\caption{(Color online). Formation of a freak wave from a
zigzag wave envelope. 
Top: at $t=0$ maximum elevation is $1.7$ m. 
Bottom: at $t=$ 16 min 39.5 s maximum elevation is $5.0$ m 
(at the zigzag corner).} 
\label{zz} 
\end{figure}

A reason why the zigzag turns are so suitable for appearance of big waves 
is the following. It is seen in Figs.\ref{Ymap_ka008_75min}-\ref{zoom} 
that the wave crests in the stripes forming zigzag pattern 
are obliquely oriented with respect to $x$ direction. Thus the group
velocity has there a non-trivial $q$-component which is negative for 
``positively'' oriented stripes (increment $\Delta x/\Delta q >0$ 
along the stripe), and positive for ``negatively'' oriented stripes.
Since the wave energy propagates basically with the group velocity,
it is accumulated near the turns where increment  $\Delta x/\Delta q$ 
changes the sign from negative to positive.
This is a genuinely 3D mechanism of rogue wave 
formation, which has not been discussed earlier in the literature.
Its additional numerical verification was performed for
non-random zigzag patterns with different initial amplitudes. It was 
found that the mechanism is only actual for sufficiently high level of
nonlinearity.
A result of evolution for nearly critical initial amplitude 
is presented in Fig.\ref{zz} 
(compare with  Figs.\ref{Ymap_ka008_75min}-\ref{zoom}).

{\it Acknowledgments.---} These investigations were supported 
by RFBR Grant 06-01-00665, by the Program ``Fundamental Problems of 
Nonlinear Dynamics'' from the RAS Presidium, 
and by Grant ``Leading Scientific Schools of Russia''.

\clearpage

\clearpage

\clearpage


\clearpage

\end{document}